\documentclass[11pt]{article}

\usepackage[margin=1in]{geometry}
\usepackage{graphicx}
\usepackage{booktabs}
\usepackage{tabularx}
\usepackage{amsmath,amssymb}
\usepackage{url}
\usepackage{placeins}
\usepackage[hidelinks]{hyperref}
\hypersetup{
  pdftitle={CAGE: Regime-Aware Algorithm Engineering for Coverage Optimization},
  pdfauthor={Amirreza Khorasanian},
  pdfsubject={Algorithm engineering for coverage optimization},
  pdfkeywords={maximum coverage, submodular optimization, algorithm engineering, Pareto analysis, bitset computation, reproducibility},
  bookmarksdepth=2
}
\title{CAGE: Regime-Aware Algorithm Engineering for Coverage Optimization}
\author{Amirreza Khorasanian\\
\small Department of Electrical and Computer Engineering, University of Tehran, Tehran, Iran\\
\small \texttt{akh5793@gmail.com}\\
\small ORCID: \href{https://orcid.org/0009-0005-2220-9854}{0009-0005-2220-9854}}
\date{}

\begin{document}
\maketitle

\begin{abstract}
Coverage-selection policies that optimize the same monotone objective can occupy different quality--runtime frontiers as workload scale, budget, repetition horizon, and execution substrate change. CAGE is a regime-aware algorithm-engineering framework that holds coverage semantics fixed through a shared bitset execution substrate and studies how the nondominated strategy set changes under controlled operating conditions. We separate intrinsic workload descriptors from measured strategy response, represent each condition by its empirical Pareto set, and identify regime transitions from changes in Pareto membership or statistically supported crossovers. Across Defects4J, Epinions, a BWSN-inspired time-to-detection workload, repeated selection, and an embedded RV32 realization, we observe distinct transitions. At a 5\% budget, one-exchange refinement adds 27 Chart lines at 11.3$\times$ baseline median runtime, whereas deeper search adds five Time lines at roughly 715$\times$. On Math, CP-SAT confirms full-universe optimality at 20\% and 40\% and residual lower-budget headroom. Persistent preprocessing reuse first achieves confidence-interval-supported speedup at horizon 20 and reaches 1.442$\times$ at horizon 100. ModelSim shows a separate bitset-width crossover. The results demonstrate that strategy nondominance is workload- and condition-dependent, supporting regime-aware algorithm engineering rather than a universal policy ranking.
\end{abstract}

\vspace{0.75em}

\section{Introduction}
Coverage selection is a recurring computational pattern: a system is given a collection of candidates, each candidate covers a subset of a universe, and only a limited subset can be selected. The limit may be a cardinality constraint, an explicit resource budget, or an operational requirement such as repeated decision making. Instantiations appear in test selection and prioritization, monitoring and sensing, graph-oriented analytics, and other domains in which the value of a candidate depends largely on the information it adds beyond what has already been selected. Coverage-guided test prioritization is a well-established example \cite{rothermel1999,rothermel2001,elbaum2001}, while sensor placement and outbreak detection provide structurally similar non-software settings \cite{leskovec2007,ostfeld2008}.

The algorithmic core is familiar: maximum coverage is a monotone submodular optimization problem with classical approximation guarantees and hardness bounds \cite{nemhauser1978,feige1998,khuller1999,sviridenko2004}. The practical ordering of implementations, however, is not determined by the objective alone. A policy that evaluates more marginal gains can obtain better coverage but incur substantial computational cost; a cheaper randomized policy can trade quality for speed; cost-aware refinement can matter only under some budgets; and reusable preprocessing can be a net loss for a single decision yet profitable over a long sequence. Low-level execution also matters because dense coverage instances repeatedly invoke set difference, union, and population-count operations. Consequently, the nondominated strategy set can change even when the coverage semantics remain fixed.

CAGE is designed to make those changes measurable. It separates selection policy from a shared bitset execution substrate: candidate coverage is represented by bitsets, marginal gain is evaluated against a common covered state, and state updates use the same bitwise semantics for every policy. The shared substrate functions as experimental control rather than as a new bitmap or approximation result: by reducing representational variation, it lets quality and execution-cost differences be attributed more cleanly to policy and operating condition.

Our central object of study is the \emph{Pareto signature} of an operating condition. We first describe a workload using intrinsic quantities that are available without running the competing strategies, including candidate and universe size, coverage-matrix density, candidate-cost dispersion when applicable, budget fraction, repeated-decision horizon, and execution substrate. Each strategy then induces a measured pair of solution quality and computational cost. The nondominated strategies form an empirical Pareto set; across a controlled sweep, the sequence of these sets is the Pareto signature. We use \emph{regime transition} for a change in this signature or for a statistically supported crossover such as the point at which reusable preprocessing becomes beneficial. Post-hoc quantities such as ``quality opportunity'' remain response variables; predictive selection for unseen instances is a separate problem left for future work.

The evaluation tests this analysis across complementary conditions rather than searching for a single winning policy. Budgeted Defects4J workloads expose saturation and expensive marginal quality; a cardinality rival study separates equal-quality implementations by execution cost; large Epinions and BWSN-inspired workloads expose explicit approximation tradeoffs; repeated selection varies the horizon over which preprocessing can be amortized; and an embedded RV32 realization varies the execution substrate and bitset width. A quality-only CP-SAT reference on Math provides an independent anchor for selected budget points: the 20\% and 40\% solutions cover the complete 1115-element universe and are solver-confirmed optimal, whereas lower budgets retain measurable headroom. These experiments therefore test both changes in Pareto membership and crossovers along controlled dimensions.

We organize the study around three research questions. \textbf{RQ1:} How does the quality--runtime Pareto signature change across workload and budget conditions under common coverage semantics? \textbf{RQ2:} At what repeated-decision horizon does persistent preprocessing reuse become beneficial? \textbf{RQ3:} How does the same bitset execution model behave when mapped to a specialized embedded substrate? Correctness and parity checks of the common abstraction are treated as experimental preconditions rather than as a separate research question.

The contributions are threefold. First, CAGE provides a shared bitset execution boundary for controlled comparison of deterministic, approximate, budget-aware, and reuse-oriented coverage-selection policies. Second, we define and apply an empirical regime-analysis method based on intrinsic workload descriptors, measured quality--runtime response, Pareto signatures, and supported crossovers; the cross-domain evaluation demonstrates budget-dependent changes in the nondominated strategy set and uses CP-SAT as an independent quality reference on Math. Third, we extend the same crossover analysis to two orthogonal dimensions---repeated-decision horizon and execution substrate---showing when persistent preprocessing and specialized bitset execution amortize their fixed overheads. Together, these contributions provide an algorithm-engineering account of \emph{when} additional optimization work remains justified across changing operating conditions.


\section{Related Work}
\subsection{Maximum coverage and submodular maximization}
The cardinality-constrained coverage objective is a canonical monotone submodular maximization problem. Classical greedy analysis establishes the familiar $1-1/e$ approximation behavior for monotone submodular set functions under cardinality constraints \cite{nemhauser1978}, while hardness results for max-$k$-cover show that this threshold is essentially unavoidable in polynomial time under standard complexity assumptions \cite{feige1998}. The budgeted variant introduces heterogeneous candidate costs. Khuller, Moss, and Naor give the classical treatment of budgeted maximum coverage \cite{khuller1999}, and Sviridenko gives a $1-1/e$ result for monotone submodular maximization under a knapsack constraint \cite{sviridenko2004}. Accordingly, our contribution lies in the algorithm-engineering question of how policies with different amounts of search effort behave once they share the same coverage representation and execution primitives, rather than in a new worst-case approximation ratio.

A second line of work reduces the computational burden of repeated marginal-gain evaluation. Minoux's accelerated greedy scheme is the historical basis for lazy evaluation of submodular gains \cite{minoux1978}. Later work studies query-efficient and threshold-style maximization \cite{badanidiyuru2014}, and stochastic greedy methods trade some solution quality for substantially fewer marginal evaluations \cite{mirzasoleiman2015}. In a sensing context, Leskovec et al. exploit submodularity and lazy evaluation for cost-effective outbreak detection in networks \cite{leskovec2007}. The CAGE evaluation includes engineering variants from these broad families and reserves published algorithm names for cases where code-level behavior supports that equivalence.

\subsection{Algorithm selection and empirical performance regimes}
The idea that different algorithms can be preferable on different instances is longstanding. Rice formalized the algorithm-selection problem as a mapping from problem features to algorithms and performance measures \cite{rice1976}; subsequent work developed per-instance algorithm portfolios and practical selector systems. SATzilla is a prominent example in which instance features are used to select among SAT solvers \cite{xu2008}, and Kotthoff surveys algorithm-selection methods for combinatorial search more broadly \cite{kotthoff2014}.

CAGE is related in motivation but different in output. Its workload descriptor separates intrinsic properties from measured responses, and controlled sweeps expose Pareto-set changes and crossover behavior. In this paper, \emph{regime-aware} therefore denotes an empirical analysis methodology; predictive per-instance selection is a natural extension rather than part of the present evaluation.

\subsection{Coverage-guided and cost-aware regression testing}
Coverage information has long been used to prioritize regression tests. Early empirical studies showed that coverage-based prioritization can improve the rate at which faults are exposed \cite{rothermel1999,rothermel2001}. Cost-cognizant work explicitly studies varying test costs and fault severities, reinforcing that selection quality and execution cost should be considered jointly \cite{elbaum2001}. Defects4J later provided a reproducible collection of real Java faults and associated test infrastructure for controlled software-testing studies \cite{just2014}. These works motivate software testing as one evaluation domain for CAGE, but the paper is not positioned as a replacement for the regression-test-prioritization literature. Here, individual JUnit test methods are candidates, production source lines are coverage elements, and isolated command runtimes provide candidate costs; this construction is used to study a general budgeted coverage objective.

\subsection{Network and sensor-placement workloads}
The Epinions workload is derived from the directed trust network distributed through SNAP, whose source citation is Richardson, Agrawal, and Domingos \cite{richardson2003}. The BWSN family originates from the Battle of the Water Sensor Networks design challenge \cite{ostfeld2008}. Our time-to-detection workload is generated from a BWSN network model and is used as a non-software temporal coverage workload rather than as a reproduction of published BWSN competition scores. These domains are useful because they exercise the same coverage-selection abstraction under very different candidate and universe structures.

\subsection{Bitset representations and execution substrates}
The common representation in CAGE is a dense bitset abstraction: marginal gains are set differences followed by population counts, and coverage updates are bitwise unions. Bitmap-oriented data structures and indexes have a long systems history; for example, Roaring bitmaps show how representation choices can materially affect set-operation performance \cite{chambi2016}. Dense bitsets serve here as an abstraction boundary shared by multiple selection policies, rather than as a new bitmap format. The embedded experiment extends this point across execution substrates by mapping the same marginal-gain and update kernels to a memory-mapped accelerator. Because the current hardware evidence is full-system ModelSim/cycle evaluation rather than post-synthesis FPGA data, it supports portability of the computational abstraction, while FPGA area, frequency, and power remain outside the measured scope.

\section{Problem Formulation}
\subsection{Coverage selection}
Let $U=\{u_1,\ldots,u_m\}$ be the coverage universe and $C=\{c_1,\ldots,c_n\}$ the candidate set. Candidate $c_i$ covers a subset $A_i\subseteq U$. For a selected set $S\subseteq C$, the coverage objective is
\begin{equation}
F(S)=\left|\bigcup_{c_i\in S} A_i\right|.
\end{equation}
We consider both cardinality constraints $|S|\le k$ and budgeted constraints
\begin{equation}
\sum_{c_i\in S} w_i\le B,
\end{equation}
where $w_i>0$ is the candidate cost.

\subsection{Bitset execution model}
Each $A_i$ is represented as a bit vector $b_i\in\{0,1\}^m$. Let $z$ denote the current covered state. The marginal gain of candidate $i$ is
\begin{equation}
\Delta(i\mid z)=\operatorname{popcount}(b_i\wedge \neg z),
\end{equation}
and selecting $i$ updates state as
\begin{equation}
z\leftarrow z\vee b_i.
\end{equation}
These operations form the shared computational interface used by the evaluated strategies.

\paragraph{Running example.}
Consider $U=\{u_1,\ldots,u_8\}$ with a cardinality limit $k=2$. Four candidates are
\[
\begin{array}{c|c|c}
\text{candidate} & A_i & b_i\;(u_1\ldots u_8)\\
\hline
c_1 & \{u_1,u_2,u_3\} & 11100000\\
c_2 & \{u_3,u_4,u_5,u_6\} & 00111100\\
c_3 & \{u_5,u_6,u_7,u_8\} & 00001111\\
c_4 & \{u_1,u_4,u_7\} & 10010010
\end{array}
\]
The engine starts from $z=00000000$. The marginal-gain kernel returns
$(3,4,4,3)$ for $(c_1,c_2,c_3,c_4)$. With deterministic candidate-id tie breaking,
a greedy policy selects $c_2$, and the update kernel produces
$z\leftarrow 00111100$. The next marginal gains are $(2,0,2,2)$; selecting $c_1$
then produces $z=11111100$ and coverage $F(S)=6$. The inputs and outputs of the
execution layer are therefore explicit: a policy supplies a candidate choice, while
the shared engine supplies gain values and the updated covered state. A naive policy
recomputes all gains at each step; a lazy policy reuses valid upper bounds; a sampled
policy evaluates only a subset; and a budget-aware policy can use the same gain/update
kernels while additionally enforcing candidate costs. The policy work changes, while
the coverage semantics remain fixed.

\subsection{Repeated decisions}
For repeated workloads, the coverage matrix is fixed while admissible candidates or other decision-state components may change over horizon $H$. A reuse method may retain static preprocessing $P$ across decisions and update only dynamic state $D_t$. The relevant cost is therefore amortized total execution over the sequence rather than one-shot latency alone.

\subsection{Intrinsic workload descriptors}
A methodological requirement is that workload descriptors not be defined using the outcome of the strategy being evaluated. We therefore write
\begin{equation}
\mathbf{x}(W)=(n,m,\delta,\nu_w,\beta,H,E),
\end{equation}
where $n$ and $m$ are candidate and universe size,
\begin{equation}
\delta=\frac{1}{nm}\sum_{i=1}^{n}|A_i|
\end{equation}
is the density of the candidate--element incidence matrix, $\nu_w=\sigma(w)/\bar w$ is the coefficient of variation of candidate cost when costs exist, $\beta=B/\sum_i w_i$ is the budget fraction for budgeted instances, $H$ is the repeated-decision horizon, and $E$ identifies the execution substrate. Not every component is active in every experiment.

\subsection{Response, Pareto signatures, and regime transitions}
For strategy $\pi$, the measured response is
\begin{equation}
\mathbf{y}_{\pi}(W)=(q_{\pi}(W),\tau_{\pi}(W)),
\end{equation}
with quality $q$ and computational cost $\tau$. Strategy $\pi_a$ Pareto-dominates $\pi_b$ when it is no worse in quality and runtime and strictly better in at least one. The empirical Pareto set is
\begin{equation}
\mathcal{P}(W)=\{\pi:\nexists\pi'\text{ that dominates }\pi\}.
\end{equation}

Many experiments vary one controlled axis $a$ while leaving the underlying coverage semantics fixed: budget fraction $\beta$, repeated-decision horizon $H$, cardinality $k$, or bitset width. For ordered settings $a_1,\ldots,a_r$, define the \emph{Pareto signature}
\begin{equation}
\Sigma_W(a_1{:}a_r)=\big(\mathcal{P}(W(a_1)),\ldots,\mathcal{P}(W(a_r))\big).
\end{equation}
We call a change in this signature an \emph{empirical regime transition}. When quality is deliberately held equal and the comparison is purely temporal, a crossover is reported only when the relevant timing ratio changes side of one; for noisy repeated measurements we additionally require the reported 95\% confidence interval to support the claimed side of break-even.

Two response-side quantities are useful for interpretation. Relative to a low-overhead baseline $\pi_f$, observed quality headroom is
\begin{equation}
h(W)=\max_{\pi}q_\pi(W)-q_{\pi_f}(W),
\end{equation}
and runtime amplification is $\rho_{\pi}=\tau_{\pi}/\tau_{\pi_f}$. These are measured outcomes, not input features. The term \emph{regime} in this paper is therefore descriptive: it refers to stable qualitative response or a transition observed along a controlled sweep, not to a learned universal classifier.

\section{The CAGE Framework}
\begin{figure}[t]
\centering
\includegraphics[width=0.92\linewidth]{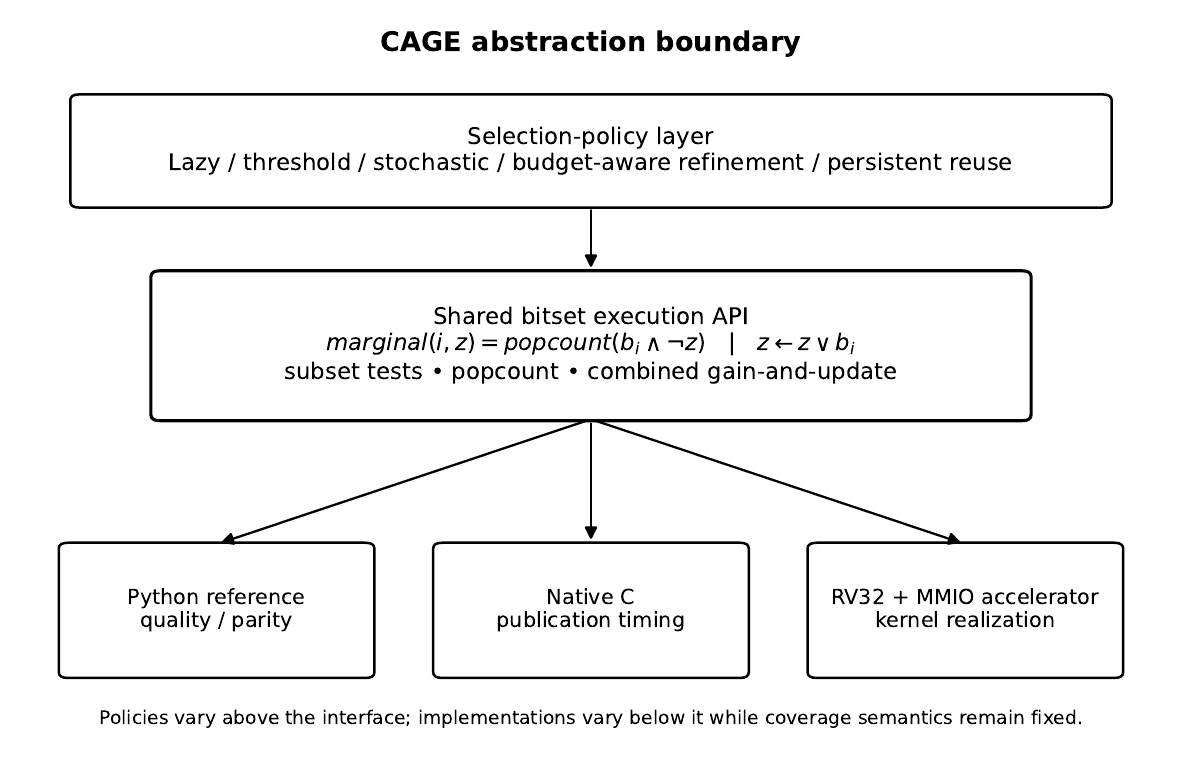}
\caption{CAGE abstraction boundary. Selection policies operate above a shared bitset API. Python provides a reference path for quality/parity, native C provides publication timing, and the RV32 realization maps a subset of the same kernels to a memory-mapped accelerator.}
\label{fig:architecture}
\end{figure}

CAGE separates selection policy from coverage execution. This boundary is the central engineering abstraction: strategies decide which candidate to select, while a shared engine owns bitset representation, marginal-gain evaluation, coverage updates, and related primitives. Figure~\ref{fig:architecture} summarizes the boundary and its three audited implementations.

\subsection{Strategy layer}
A selection strategy is modeled as a policy
\begin{equation}
\pi:(S,C_{\mathrm{active}})\rightarrow c_i,
\end{equation}
where $S$ contains the current optimization state and $C_{\mathrm{active}}$ is the admissible candidate set. Policies may use exact marginal gains, lazy upper bounds, thresholds, randomized sampling, candidate costs, local refinements, or reusable incidence information. Their asymptotic behavior may differ; the invariant is that they consume the same coverage semantics.

At a high level, every policy executes the same state transition: initialize $z=0$ and $S=\emptyset$; while the policy can return a feasible candidate $i$, append $i$ to $S$, update $z\leftarrow z\vee b_i$, and expose the new state to the policy. In the running example from Section~3, the engine therefore returns the same gain values and covered-state update regardless of whether the policy reaches $c_2$ by full scanning, lazy bounds, sampling, or a cost-aware rule. Policy-specific work determines how $i$ is chosen; the meaning of gain and coverage is invariant.

\subsection{Bitset execution engine}
The execution layer stores each candidate as a bitset and implements the recurring primitives required by selection. The core pair is marginal gain, $\operatorname{popcount}(b_i\wedge\neg z)$, and state update, $z\leftarrow z\vee b_i$. Additional kernels include subset tests, raw population count, and a combined gain-and-update operation. This makes representation an abstraction boundary rather than a property of one selection policy.

\subsection{Strategy families used in the evaluation}
The experiments intentionally span different policy behaviors. The cardinality study includes deterministic lazy and naive baselines, threshold/incidence engineering variants, and randomized stochastic/sampled-lazy variants (internal code label LTLG). The budget-tier study uses four explicitly ordered levels of additional search effort. \texttt{fast} is native lazy ratio-greedy, scoring feasible candidates by marginal gain per cost and retaining the best feasible singleton as a safeguard. \texttt{practical} starts from \texttt{fast} and iterates feasible additions and one-out/one-in improvements until no lexicographically tie-broken quality improvement remains. \texttt{balanced} compares this result with an alternate lazy seed based on $\Delta/\sqrt{w}$ followed by the same one-exchange refinement, returning the better solution. \texttt{deep} starts from \texttt{balanced} and adds bounded two-exchange search over a diversified candidate pool of at most 12 candidates. The reuse study contrasts ordinary lazy execution with persistent compressed-incidence preprocessing. Table~\ref{tab:strategies} summarizes the roles without implying that every listed strategy is algorithmically novel.

\begin{table*}[t]
\centering
\footnotesize
\caption{Strategy roles in the audited experiments. Budget-tier definitions are stated at implementation level, and family labels distinguish established patterns from CAGE engineering variants.}
\label{tab:strategies}
\begin{tabularx}{\textwidth}{@{}p{0.22\textwidth} p{0.31\textwidth} X@{}}
\toprule
Strategy / family & Audited role or definition & Main tradeoff exposed \\
\midrule
\texttt{lazy\_reuse} & Deterministic lazy-greedy baseline & Greedy quality with reduced marginal re-evaluation \\
\texttt{v45\_naive} & Deterministic direct-evaluation reference & Cost of recomputing marginal gains \\
Threshold / incidence variants & Deterministic engineering variants & Evaluation work vs. policy fidelity / indexing overhead \\
Stochastic / sampled-lazy variants (internal code label LTLG) & Randomized approximate variants & Quality vs. runtime \\
\texttt{fast} & Lazy $\Delta/w$ greedy + best feasible singleton & Low-overhead budgeted selection \\
\texttt{practical} & \texttt{fast} + iterative add / one-out-one-in improvement & Extra quality for local-search cost \\
\texttt{balanced} & Best of \texttt{practical} and $\Delta/\sqrt{w}$ seed + one-exchange & Alternative cost weighting + refinement \\
\texttt{deep} & \texttt{balanced} + bounded two-exchange over pool $\le12$ & Higher search effort for remaining headroom \\
\texttt{compressed\_incidence} & Persistent preprocessed incidence method & Preprocessing cost vs. repeated-decision amortization \\
\bottomrule
\end{tabularx}
\end{table*}


\paragraph{Method lineage.}
The deterministic lazy baseline uses the standard diminishing-return upper-bound idea associated with accelerated greedy evaluation \cite{minoux1978}, while the naive reference recomputes every active marginal gain. The randomized stochastic variant uses a stochastic-greedy-style sample-size schedule \cite{mirzasoleiman2015} but samples from the remaining active pool; the LTLG implementation further orders that sampled pool by cached upper gains and may stop evaluating it once the bound cannot beat the current best. The threshold variant uses a multiplicatively decreasing gain threshold in the spirit of threshold/query-efficient submodular maximization \cite{badanidiyuru2014}. These implementation details motivate the family labels, while the manuscript describes them conservatively as engineering variants rather than exact reproductions of the published algorithms. The incidence structures and budget-tier local refinements are likewise presented as engineering variants; approximation guarantees remain those of the established methods cited above where applicable.

\subsection{Regime-analysis pipeline}
\begin{figure}[t]
\centering
\includegraphics[width=0.96\linewidth]{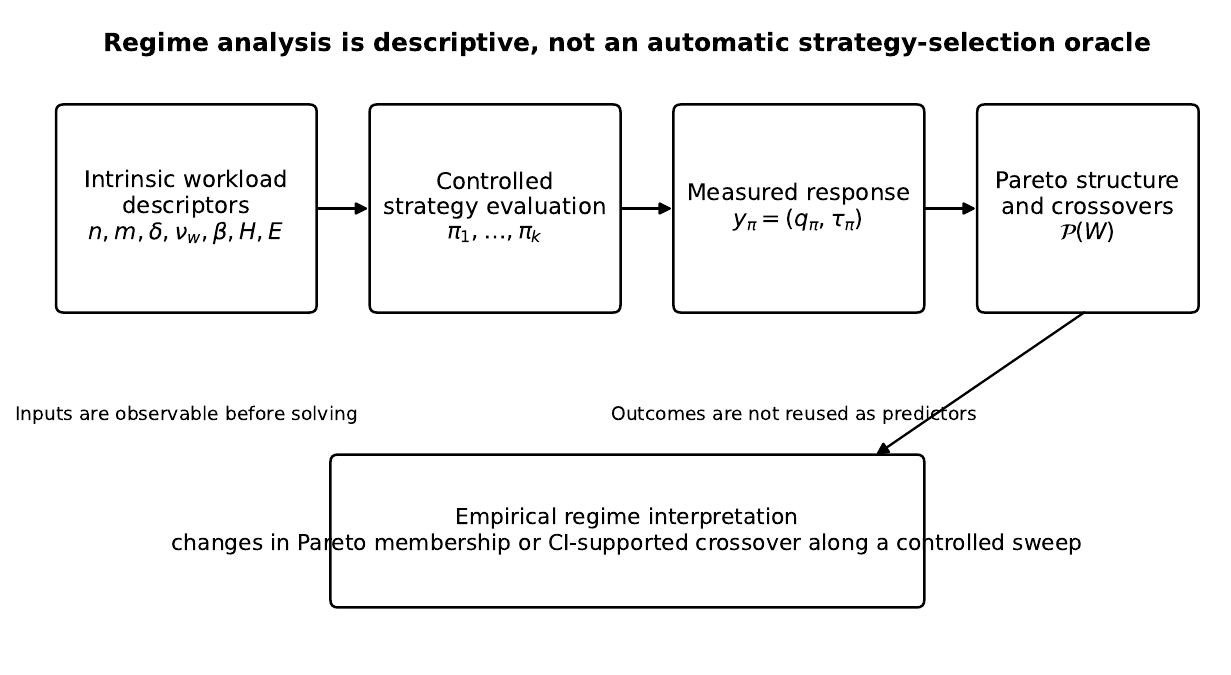}
\caption{Regime-analysis pipeline. Observable workload descriptors and controlled sweep variables are kept separate from measured quality--runtime response. Regime statements are based on Pareto-set changes or supported crossovers, not on an unvalidated strategy-selection oracle.}
\label{fig:regime-pipeline}
\end{figure}

The framework records intrinsic descriptors separately from outcomes. Structural descriptors organize the experiment; controlled sweeps vary one axis; measured quality and runtime define the Pareto set; and the resulting Pareto signature or supported crossover is interpreted as the empirical regime evidence. Figure~\ref{fig:regime-pipeline} makes this data flow explicit. In particular, response quantities such as ``remaining quality headroom'' are never fed back as if they were known a priori.

\subsection{Cross-layer execution}
Because the strategy layer communicates through the bitset abstraction, the execution primitives can be realized on different substrates. In the software experiments they are implemented in native C for timing and mirrored by deterministic Python quality evaluation. The embedded realization maps marginal-gain, gain-and-update, subset, covered-state, and popcount operations to a memory-mapped accelerator attached to a PicoRV32 RV32IM system. The hardware experiment tests transfer of the computational abstraction and locates the overhead crossover; post-synthesis FPGA metrics remain outside the current evidence.

\section{Experimental Methodology}
\subsection{Final evidence set and workload selection}
Headline timing, Pareto, and crossover claims use the final audited runs recorded in the reproducibility manifest; earlier development runs are excluded from the reported analysis. A separate quality-only CP-SAT reference is retained for the Math budget workload to anchor solution headroom. Its four budget records use the same prepared workload semantics as the final tier experiment: the tier-method quality rows agree exactly on record name, budget, coverage, selected count, total cost, and feasibility. CP-SAT uses one search worker and integer cost scaling of $10^6$; solver status therefore applies to this scaled integer-cost model, and solver time is not compared with native C selection timing. At 20\% and 40\%, full-universe coverage also supplies an absolute quality upper bound independent of solver timing. Table~\ref{tab:workloads} summarizes the primary workloads. Defects4J supplies real Java revisions and test infrastructure \cite{just2014}; Epinions is derived from the directed trust network distributed by SNAP \cite{richardson2003}; and the BWSN-inspired workload is generated from a water-network model related to the BWSN challenge \cite{ostfeld2008}.

\begin{table*}[t]
\centering
\footnotesize
\caption{Primary workloads and the controlled axis used to expose quality--cost behavior.}
\label{tab:workloads}
\begin{tabularx}{\textwidth}{@{}>{\raggedright\arraybackslash}p{0.17\textwidth} >{\raggedright\arraybackslash}p{0.15\textwidth} r r >{\raggedright\arraybackslash}p{0.15\textwidth} >{\raggedright\arraybackslash}X@{}}
\toprule
Workload & Domain & $n$ & $m$ & Controlled axis & Primary role \\
\midrule
Math 1f & Defects4J & 313 & 1,115 & $k$; budget $\beta$ & Cardinality rivals and budget transition \\
Chart 1f & Defects4J & 436 & 4,579 & budget $\beta$ & Expensive marginal quality \\
Time 1f & Defects4J & 4,011 & 3,643 & budget $\beta$ & Large candidate-set overhead \\
Mockito 1f & Defects4J & 978 & 2,617 & budget $\beta$ & Budget-dependent Pareto transition \\
Epinions 10k & Social network & 10,000 & 75,879 & $k$ & Randomized quality--speed tradeoff \\
BWSN TTD & Sensor/time & 126 & 2,898 & $k$ & Non-software temporal coverage \\
Math reuse (record 2) & Repeated selection & 313 & 1,115 & horizon $H$ & Preprocessing amortization \\
Embedded RV32 & Execution substrate & -- & -- & bitset width / $E$ & Kernel overhead crossover \\
\bottomrule
\end{tabularx}
\end{table*}


The Defects4J formulation treats individual JUnit test methods as candidates, covered production source lines as the universe, and median isolated test-command wall-clock time as candidate cost. The evaluated budget fractions are $\beta\in\{0.05,0.10,0.20,0.40\}$ of measured full-suite candidate cost. Coverage generation and workload preparation are not included in native selection timing. Table~\ref{tab:d4j-descriptors} reports structural descriptors computed directly from the prepared masks and cost sidecars. They are used to contextualize results, not to fit a predictive model.

\begin{table}[t]
\centering
\small
\caption{Audited intrinsic descriptors for the four budgeted Defects4J workloads. $\delta$ is candidate--element matrix density and $\nu_w$ is the coefficient of variation of candidate cost.}
\label{tab:d4j-descriptors}
\begin{tabular}{@{}lrrrr@{}}
\toprule
Workload & $n$ & $m$ & $\delta$ & $\nu_w$ \\
\midrule
Math 1f & 313 & 1,115 & 0.258 & 0.217 \\
Chart 1f & 436 & 4,579 & 0.129 & 0.059 \\
Time 1f & 4,011 & 3,643 & 0.275 & 0.025 \\
Mockito 1f & 978 & 2,617 & 0.269 & 0.519 \\
\bottomrule
\end{tabular}
\end{table}


\subsection{Separation of quality and timing}
Quality and timing are deliberately separated. Deterministic Python evaluation establishes reference quality rows for budget-tier experiments. Native C timing is performed separately, and native outputs are parity-checked against coverage, selected count, and total cost. For randomized experiments with separated quality and timing phases, independent reseeded trials estimate quality while their timing is discarded; publication timing is obtained from isolated paired blocks. This prevents parallel quality throughput from contaminating timing claims.

\subsection{Timing protocols}
All primary budget-tier runs use MSVC Release builds, timing on logical CPU 1, a 250 ms cooldown, and randomized cyclic near-counterbalanced four-method timing blocks. Math and Chart use 21 blocks with 200 timed repetitions and 20 warmups. Time and Mockito retain 21 timing blocks but use 10 repetitions and 2 warmups because the full inner repetition count was prohibitive. The unit of statistical replication is the timing block, so this reduction lowers within-block averaging while retaining all 21 outer block observations.

Epinions and BWSN TTD separated-phase runs use 200 independent reseeded quality trials and 21 paired timing blocks with 200 repetitions and 20 warmups on CPU 1. The reuse experiment uses five candidate-availability sequences, seven outer timing runs, warmup 2, randomized method order, and equality checks at every tested horizon. Table~\ref{tab:protocols} summarizes the protocols.

\begin{table*}[t]
\centering
\small
\caption{Publication timing and quality protocols for the primary audited runs.}
\label{tab:protocols}
\begin{tabularx}{\textwidth}{@{}p{0.20\textwidth} p{0.25\textwidth} p{0.25\textwidth} X@{}}
\toprule
Experiment & Quality protocol & Timing protocol & Note \\
\midrule
Math/Chart budget & Deterministic Python reference & 21 blocks $\times$ 200 reps; 20 warmups & CPU 1, 250 ms cooldown \\
Time/Mockito budget & Deterministic Python reference & 21 blocks $\times$ 10 reps; 2 warmups & Reduced inner reps due prohibitive runtime \\
Epinions split quality/timing & 200 independent reseeded trials & 21 blocks $\times$ 200 reps; 20 warmups & Quality timing discarded \\
BWSN TTD split quality/timing & 200 independent reseeded trials & 21 blocks $\times$ 200 reps; 20 warmups & Quality timing discarded \\
Reuse & Equality checked at every horizon & 5 sequences $\times$ 7 outer runs & Randomized method order; warmup 2 \\
Embedded & SW/HW result equality in firmware & Full-system processor cycles & ModelSim, not post-synthesis FPGA \\
\bottomrule
\end{tabularx}
\end{table*}


\subsection{Use of generative AI}
During manuscript preparation, the author used OpenAI ChatGPT to assist with manuscript organization and language revision, code and documentation review, and preparation of reproducibility and plotting scripts. The tool was not treated as an author or as an autonomous source of scientific evidence. All algorithms, experimental data, numerical results, citations, interpretations, and final manuscript content were reviewed and verified by the author, who takes full responsibility for the work.

\subsection{Metrics and statistical summaries}
For deterministic budget experiments we report coverage, normalized coverage, median native runtime, bootstrap confidence intervals produced by the analysis pipeline, and empirical Pareto membership. For randomized split experiments we report mean coverage relative to the lazy baseline and paired timing-block speedup summaries. For reuse we report total speedup over the repeated-decision horizon and distinguish the first median win from the first 95\% confidence-interval-supported win. Embedded results use processor cycles measured by the full-system cycle counter. These metrics instantiate the response vector $\mathbf{y}_{\pi}(W)$ and Pareto signatures defined in Section~3.

\section{Evaluation}
Before answering the research questions, we validate the common abstraction. The final experiments execute deterministic baselines, threshold/incidence variants, randomized approximate methods, budget-aware refinements, and persistent-reuse variants over the same coverage semantics. In the budget-tier pipeline, native outputs are parity-checked against deterministic Python quality rows before summary generation. In the reuse experiment, coverage, selected candidates, and active sequences match at every tested horizon and the validation-failure file is empty. These checks are experimental preconditions: they hold the coverage definition fixed so that subsequent differences trace to policy and execution choices.

\subsection{RQ1: How does the quality--runtime Pareto structure change across workload conditions?}
The budget sweeps provide the cleanest regime evidence because the underlying prepared workload remains fixed while $\beta$ changes. Table~\ref{tab:pareto-signatures} reports the resulting Pareto signatures. Every audited Defects4J workload changes Pareto membership somewhere in the sweep, and all four collapse to \texttt{fast} alone by $\beta=0.20$. The transition point and the methods that remain competitive before saturation differ by workload.

\begin{table*}[t]
\centering
\small
\caption{Budget-sweep Pareto signatures for the audited Defects4J tier experiment. Sets contain the nondominated methods under measured coverage and median native runtime. $F$, $P$, $B$, and $D$ denote \texttt{fast}, \texttt{practical}, \texttt{balanced}, and \texttt{deep}. A change across columns is an empirical regime transition under the definition in Section~3.}
\label{tab:pareto-signatures}
\begin{tabular}{@{}lcccc@{}}
\toprule
Workload & $\beta=0.05$ & $\beta=0.10$ & $\beta=0.20$ & $\beta=0.40$ \\
\midrule
Math & $\{F,P\}$ & $\{F,B\}$ & $\{F\}$ & $\{F\}$ \\
Chart & $\{F,P\}$ & $\{F,P\}$ & $\{F\}$ & $\{F\}$ \\
Time & $\{F,P,B\}$ & $\{F\}$ & $\{F\}$ & $\{F\}$ \\
Mockito & $\{F,P\}$ & $\{F,P,B\}$ & $\{F\}$ & $\{F\}$ \\
\bottomrule
\end{tabular}
\end{table*}


\paragraph{Math separates cardinality behavior from budget behavior.}
The Math cardinality-rival study shows that several deterministic strategies can attain identical coverage while differing substantially in runtime; randomized variants then expose an explicit quality--speed tradeoff. The separate budget experiment has small headroom among the tier methods at $\beta=0.05$ (1034 lines for \texttt{fast} versus 1036 for \texttt{practical}), changes Pareto membership again at $\beta=0.10$, and reaches 1115/1115 coverage for every tier method by $\beta=0.20$. A quality-only CP-SAT reference sharpens this interpretation (Table~\ref{tab:math-cpsat}): at 5\%, its feasible incumbent covers 1039 lines but optimality is not proved; at 10\%, the scaled formulation is solved to optimality at 1086, one line above the best tier method; and at 20\% and 40\%, 1115 is proven optimal. Thus the low-budget tier frontier does not exhaust all quality headroom, whereas the higher-budget saturation is solver-confirmed. Accordingly, Math is interpreted conditionally by formulation and budget rather than through one unconditional regime label.

\begin{figure}[t]
\centering
\includegraphics[width=0.80\linewidth]{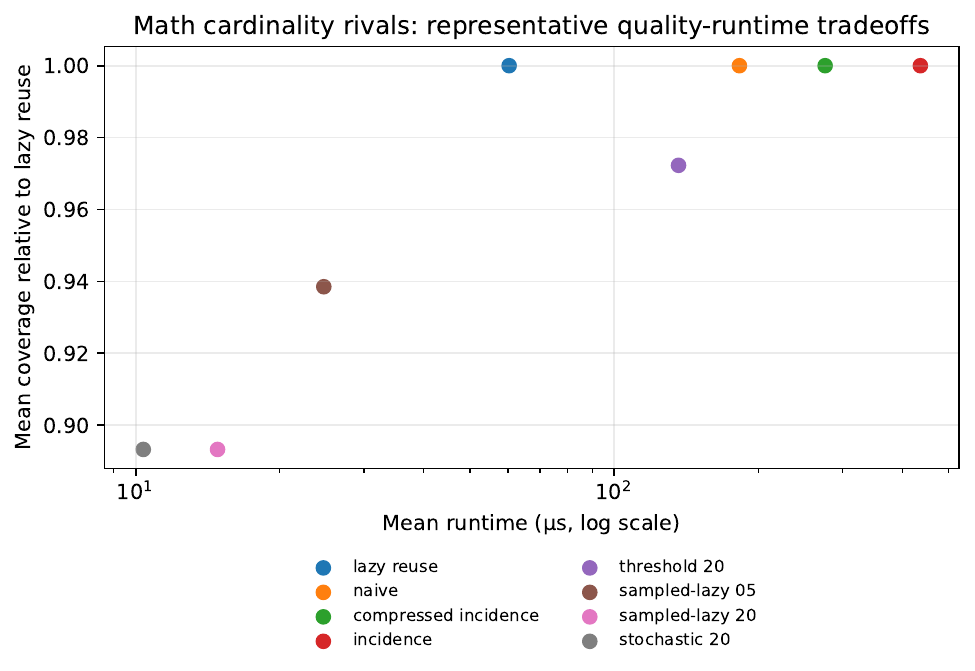}
\caption{Representative methods from the Math cardinality-rival study. Deterministic equal-quality points and randomized quality--speed tradeoffs are distinct from the separate Math budget sweep.}
\label{fig:math-rivals}
\end{figure}

\begin{table}[t]
\centering
\small
\caption{Quality-only CP-SAT reference for the Math budget workload. CP-SAT runtime is not compared with native selection timing. At 5\%, the solver returned a feasible incumbent but did not prove optimality within the configured limit.}
\label{tab:math-cpsat}
\begin{tabular}{@{}rrrrr@{}}
\toprule
$\beta$ & Best tier & CP-SAT & Status & Best bound \\
\midrule
0.05 & 1036 & 1039 & FEASIBLE & 1069 \\
0.10 & 1085 & 1086 & OPTIMAL & 1086 \\
0.20 & 1115 & 1115 & OPTIMAL & 1115 \\
0.40 & 1115 & 1115 & OPTIMAL & 1115 \\
\bottomrule
\end{tabular}
\end{table}


\paragraph{Chart, Time, and Mockito exhibit different low-budget Pareto signatures.}
At $\beta=0.05$, Chart places \texttt{fast} and \texttt{practical} on the frontier: 4339 lines at approximately 0.1912 ms versus 4366 lines at approximately 2.1644 ms. The 27-line improvement costs roughly an 11.3$\times$ ratio of reported medians. The Pareto set remains $\{F,P\}$ at $\beta=0.10$ and collapses to $\{F\}$ after full coverage is reached at $\beta=0.20$.

Time is more extreme. At $\beta=0.05$, \texttt{fast}, \texttt{practical}, and \texttt{balanced} are nondominated, but the marginal return is tiny relative to runtime: \texttt{fast} covers 3609 lines in approximately 3.18 ms, whereas \texttt{deep} covers 3614 in approximately 2.274 s. At $\beta=0.10$, all methods reach the full 3643-line universe and only \texttt{fast} remains Pareto-optimal. This is consistent with a scale/overhead-limited condition rather than a claim that deeper search can never improve Time.

Mockito has a different signature. At $\beta=0.05$, \texttt{fast} and \texttt{practical} are nondominated; at $\beta=0.10$, \texttt{balanced} also enters the Pareto set by gaining one additional line over \texttt{practical}; and at $\beta\ge0.20$, all methods cover 2617/2617 and only \texttt{fast} remains nondominated. At $\beta=0.20$, \texttt{deep} uses 367.819 ms versus 0.3554 ms for \texttt{fast}, a roughly $10^3$-fold ratio of reported medians with no coverage benefit.

\begin{figure*}[t]
\centering
\begin{minipage}[t]{0.48\textwidth}\centering
\includegraphics[width=\linewidth]{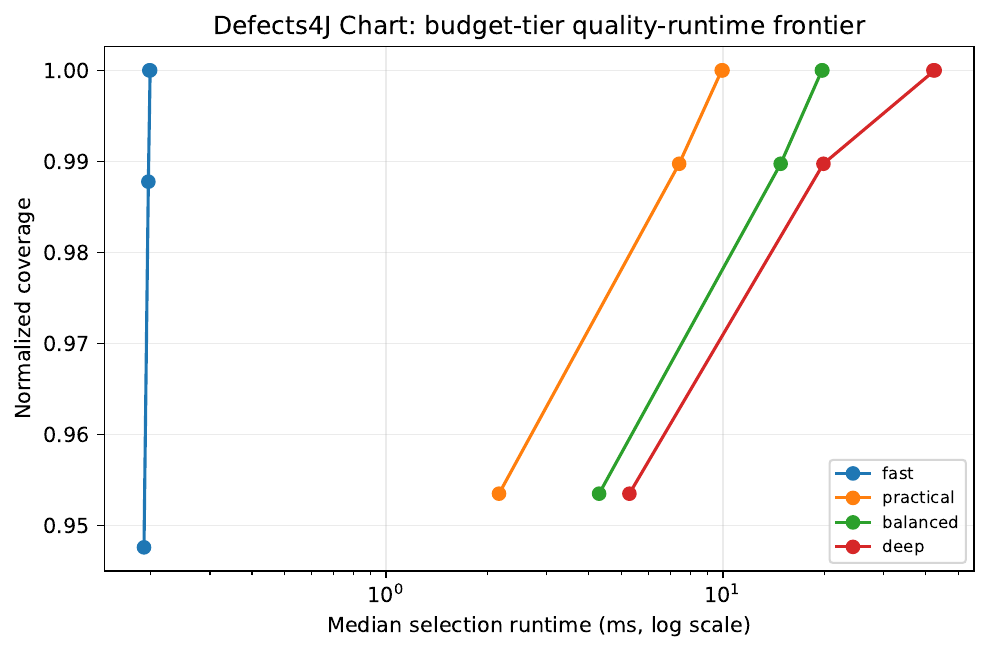}\\[-2pt]
\textbf{(a) Chart}
\end{minipage}\hfill
\begin{minipage}[t]{0.48\textwidth}\centering
\includegraphics[width=\linewidth]{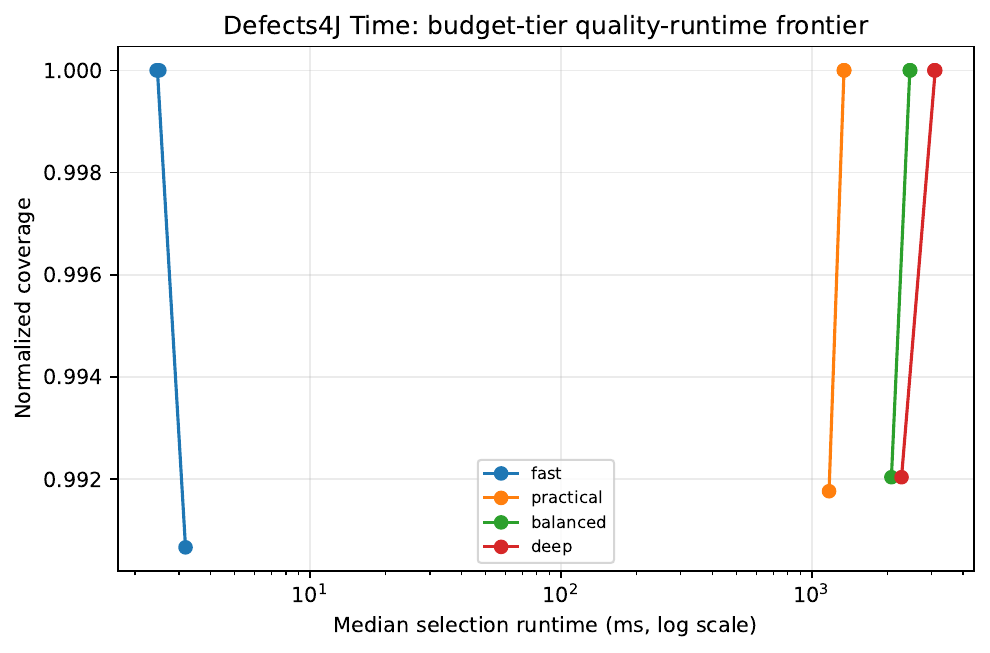}\\[-2pt]
\textbf{(b) Time}
\end{minipage}

\vspace{3pt}
\begin{minipage}[t]{0.48\textwidth}\centering
\includegraphics[width=\linewidth]{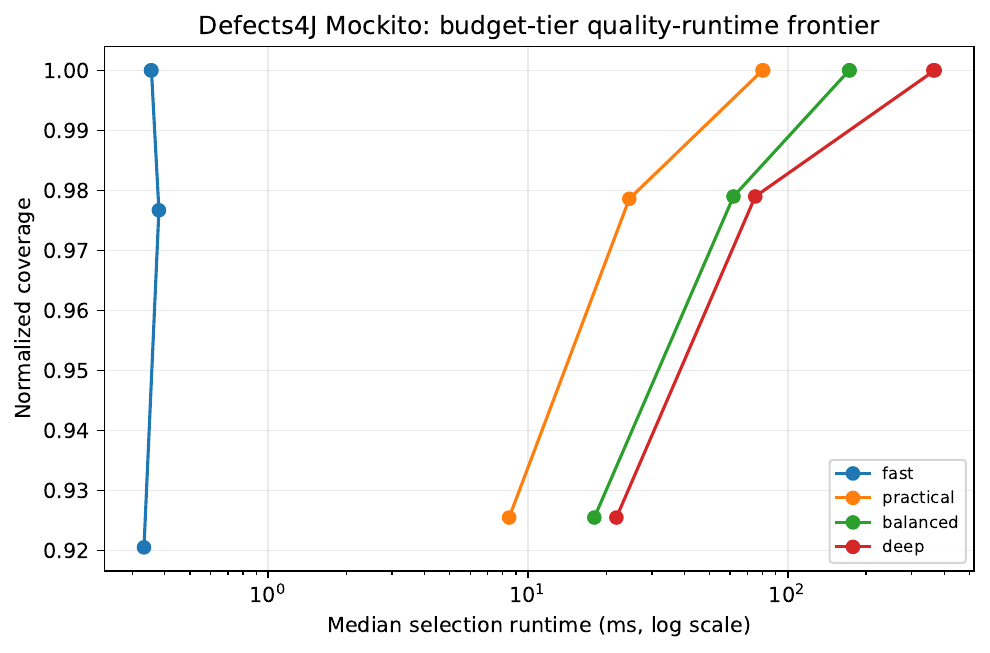}\\[-2pt]
\textbf{(c) Mockito}
\end{minipage}
\caption{Defects4J budget-tier quality--runtime frontiers. Larger panels expose distinct low-budget Pareto structures and the eventual collapse to the low-overhead baseline after coverage saturation.}
\label{fig:d4j-frontiers}
\end{figure*}

\paragraph{Large and non-software workloads expose approximation tradeoffs.}
The Epinions separated quality/timing experiment provides a 10,000-candidate, 75,879-element workload. At $k=500$, sampled-lazy 20 has mean coverage ratio approximately 0.8813 relative to lazy selection and median speedup approximately 1.2015$\times$; sampled-lazy 05 retains approximately 0.9535 of baseline coverage with about 1.0586$\times$ speedup. The BWSN-inspired TTD workload exhibits a stronger exchange: at $k=10$, sampled-lazy 20 reaches approximately 0.9111 mean coverage ratio with about 1.6947$\times$ median speedup, while sampled-lazy 05 reaches approximately 0.9495 with about 1.2336$\times$ speedup. The BWSN experiment is a generated time-layer workload used for cross-domain temporal coverage analysis, rather than a reproduction of published BWSN competition numbers.

\begin{figure*}[t]
\centering
\begin{minipage}[t]{0.49\textwidth}\centering
\includegraphics[width=\linewidth]{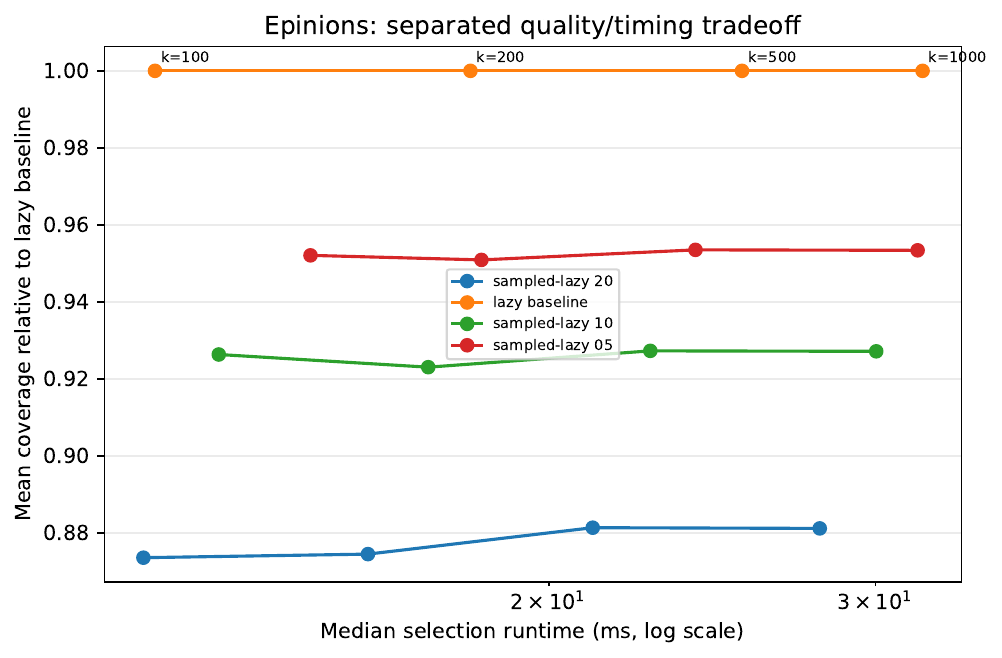}\\[-2pt]
\textbf{(a) Epinions}
\end{minipage}\hfill
\begin{minipage}[t]{0.49\textwidth}\centering
\includegraphics[width=\linewidth]{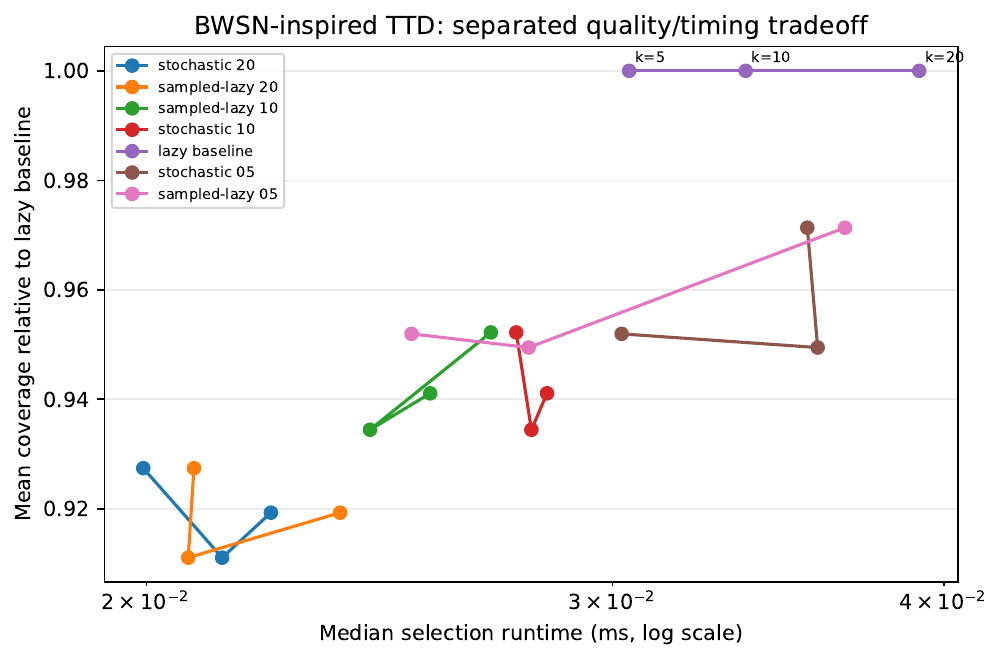}\\[-2pt]
\textbf{(b) BWSN-inspired TTD}
\end{minipage}
\caption{Randomized quality--runtime tradeoffs outside the budgeted Defects4J setting. Quality and timing are measured in separate phases.}
\label{fig:approx-frontiers}
\end{figure*}

\subsection{RQ2: When does persistent preprocessing reuse pay off?}
The repeated-decision experiment fixes the Math coverage matrix ($n=313$, $m=1115$, $k=63$) and applies deterministic candidate-availability drift across horizons $H\in\{1,5,10,20,100\}$. Coverage, selected candidates, and active sequences are identical between the lazy baseline and the compressed-incidence method, isolating amortized execution cost rather than a quality tradeoff.

Reuse exhibits an amortization crossover. The first median total-speedup win occurs at $H=10$, and the first tested horizon whose 95\% confidence interval lies fully above one is $H=20$. At $H=100$, the median total speedup is 1.441964$\times$ with 95\% CI [1.411861, 1.448990]. Under the definition in Section~3, $H=20$ is the first confidence-supported temporal crossover in the tested sweep.

\begin{figure}[t]
\centering
\includegraphics[width=0.74\linewidth]{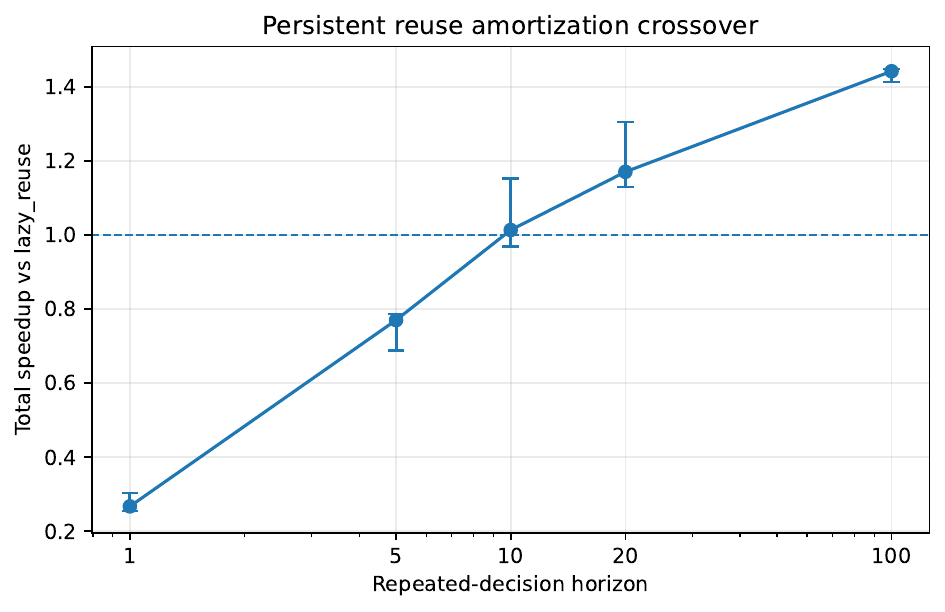}
\caption{Persistent-reuse amortization. The horizontal line at speedup 1 marks break-even; error bars show the reported 95\% confidence intervals.}
\label{fig:reuse}
\end{figure}

\subsection{RQ3: How does the bitset execution model behave on an embedded substrate?}
The embedded realization integrates a memory-mapped bitset accelerator with a PicoRV32 RV32IM system in full-system ModelSim simulation. Firmware boots from instruction memory, exercises the accelerator through MMIO, and compares software-only and hardware-assisted selected IDs and coverage before reporting pass status. The final run reports \texttt{FULL\_SYSTEM\_PASS} at 1,473,018 cycles with zero ModelSim errors.

The cycle comparison shows a setup-overhead crossover. At one 32-bit word the hardware-assisted path is slower (0.877$\times$), and at two words it is approximately break-even (0.998$\times$). At four words the speedup is 1.219$\times$, increasing to 2.735$\times$ at 32 words. The evidence therefore supports transfer of the bitset kernels and a width-dependent cycle crossover. FPGA area, frequency, power, and post-synthesis performance remain outside the measured scope.

\begin{figure}[t]
\centering
\includegraphics[width=0.72\linewidth]{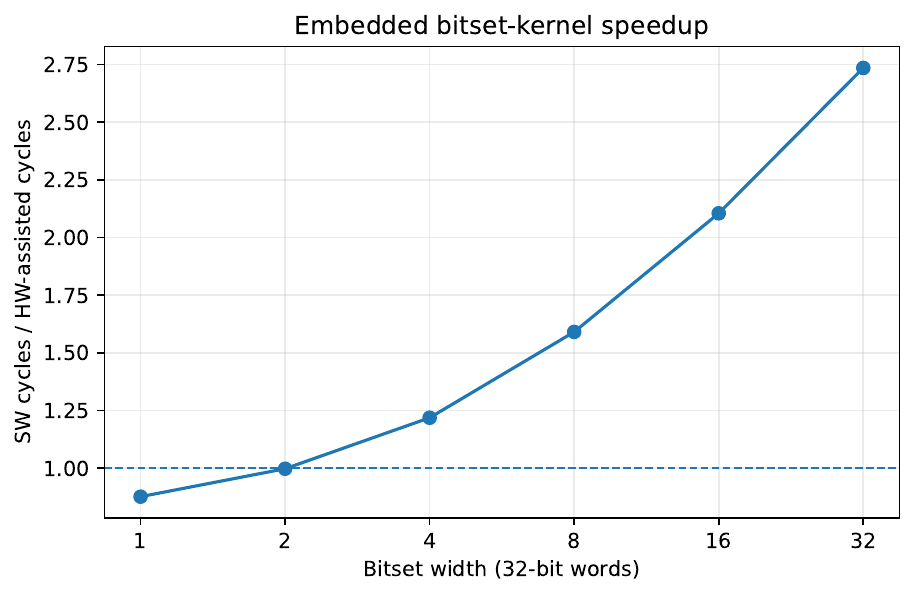}
\caption{Full-system embedded cycle speedup versus bitset width. Results are ModelSim/cycle measurements, not post-synthesis FPGA implementation results.}
\label{fig:hardware}
\end{figure}

\subsection{Evaluation summary}
The central empirical object is the changing Pareto signature. Budget sweeps alter which refinement levels are nondominated; cardinality and randomized experiments expose equal-quality efficiency differences and quality--speed exchanges; repeated execution yields a confidence-supported horizon crossover; and the embedded realization yields a width crossover. These are the empirical bases for the workload-regime interpretation of CAGE.

\FloatBarrier


\section{Discussion}
The evaluation shifts the emphasis from algorithm competition to algorithm engineering. When policies share a common coverage abstraction, the practical question is not simply which policy can extract the most coverage, but whether the workload and operating point leave enough exploitable headroom to justify the additional work. The Pareto signatures make this statement concrete: for each audited Defects4J workload, the set of nondominated tier methods changes as the budget changes, and by $\beta=0.20$ the low-overhead baseline alone remains nondominated.

Intrinsic descriptors provide context rather than causal identification. Time has the largest Defects4J candidate set ($n=4011$) and very low candidate-cost dispersion ($\nu_w\approx0.025$); its low-budget quality premium is only a few lines while local refinement costs orders of magnitude more. Mockito has a smaller candidate set but much higher cost dispersion ($\nu_w\approx0.519$), and its Pareto membership changes between the 5\% and 10\% budgets before saturation. Chart has lower incidence density and low cost dispersion yet retains a measurable low-budget quality premium. These contrasts motivate regime analysis; the current cross-workload evidence supports association together with controlled within-workload sweeps, while causal modeling remains a separate research problem.

The Math CP-SAT reference adds a complementary quality anchor. It shows that heuristic agreement alone is insufficient evidence of optimality: at low budgets the solver finds additional coverage, whereas at 20\% and 40\% the full 1115-element universe is attained. We use this reference to interpret solution headroom, with solver time kept separate from native timing; broader exact-solver certification remains future work.

A second implication is that representation can serve as an abstraction boundary. Bitset coverage semantics remain fixed while policies vary above them and implementations vary below them. This permits separate reasoning about selection policy, persistent-state reuse, and specialized execution. The embedded experiment is useful for this reason even though it is intentionally limited to simulation: it tests whether the same kernel interface survives a substrate change and where fixed interface overhead is amortized.

Finally, ``regime-aware'' remains an empirical claim. The present work records observable descriptors, performs controlled sweeps, and reports Pareto-set changes and supported crossovers. A natural extension is an adaptive selector that estimates structural descriptors and chooses a policy under a user-specified quality or latency objective; that predictive problem requires its own training and validation protocol.

\section{Threats to Validity}
\paragraph{Workload coverage and causal interpretation.} The evaluation spans several domains and scales, but it is not exhaustive. Controlled sweeps support within-workload transition claims; associations between structural descriptors and behavior across different workloads do not establish that a descriptor alone causes the observed regime. Regime labels therefore describe evaluated configurations rather than a universal taxonomy.

\paragraph{Strategy coverage and method lineage.} The experiments span deterministic, approximate, budget-aware, and reuse-oriented policies but do not contain every known maximum-coverage or submodular-optimization method. Some implementation variants are engineering implementations inspired by broad algorithmic families rather than exact reproductions of a particular publication. The paper therefore avoids attribution of equivalence unless supported by code-level mapping.

\paragraph{Timing environment.} Native timing is tied to the audited MSVC/Windows environment and CPU-affinity protocol. Absolute timings may change on other processors, compilers, and operating systems. We emphasize paired comparisons, Pareto membership, and crossovers rather than hardware-independent latency constants. Time and Mockito use fewer inner repetitions than Math and Chart because the full protocol was prohibitive; all retain 21 timing blocks and the same isolation/order design.

\paragraph{Defects4J cost semantics.} Candidate cost is the median isolated Defects4J test-command wall-clock time on the fixed revision. It therefore includes command/JVM/build overhead and is not identical to in-process unit-test execution cost. The relevant-test scope targets selection for a fixed revision and should not be interpreted as whole-project test-suite minimization.

\paragraph{Quality-reference scope.} CP-SAT is used only as a quality reference for the Math budget workload. Its optimality status refers to the integer-cost formulation described in Methodology, and its solver runtime is not compared with native C timing. The absence of an exact reference on the other workloads means that claims there are about observed Pareto structure and saturation within the evaluated strategy set unless full-universe coverage supplies an absolute upper bound.

\paragraph{Representation scope.} The software comparisons intentionally hold a dense bitset representation fixed. This isolates policy behavior but does not establish that dense bitsets dominate sparse or compressed representations on every workload. Alternative representation choices are therefore outside the present comparison.

\paragraph{BWSN scope.} The time-to-detection experiment uses a generated BWSN-inspired ensemble. It is a cross-domain temporal coverage workload, not a reproduction of published BWSN competition results.

\paragraph{Hardware scope.} Embedded evidence is full-system ModelSim/cycle-level evaluation. No claims are made about FPGA resource utilization, frequency closure, power, post-synthesis latency, or end-to-end application acceleration.

\section{Conclusion}
CAGE studies coverage selection by holding coverage semantics and the bitset execution boundary fixed while varying policy and operating condition. The resulting analysis uses intrinsic workload descriptors, measured quality--runtime response, Pareto signatures, and supported crossovers to characterize when additional optimization work changes the nondominated strategy set.

The experiments show that this set is not stable across conditions. Low-budget refinement can buy measurable coverage, the same extra search can become effectively pure overhead after saturation, approximation can exchange quality for speed on larger workloads, persistent preprocessing becomes worthwhile only after enough repeated decisions, and specialized bitset execution becomes favorable only after fixed interface costs are amortized. The Math CP-SAT reference further shows why independent quality anchors matter: higher-budget saturation is solver-confirmed, while lower budgets still leave residual headroom.

The central conclusion is therefore about operating regimes rather than algorithm ranking. For coverage-selection systems, the relevant engineering question is not simply which policy is strongest in isolation, but which policies remain nondominated under the workload scale, budget, temporal horizon, and execution environment at hand. The present work establishes an empirical methodology for answering that question. A natural next step is a separately validated selector that predicts useful policies from intrinsic descriptors and user-specified quality or latency objectives; such prediction is intentionally outside the claims made here.


\appendix

\section{Extended Experimental Details}
The accompanying reproducibility manifest records exact final run directories, method sets, timing repetitions, warmups, CPU assignments, cooldowns, and hashes; the frozen v1.0.0 research artifact is archived at Zenodo \cite{cage_artifact2026}. Machine-readable per-budget and per-$k$ summaries are distributed with the evidence package so that the manuscript's condensed tables can be regenerated without relying on transcribed numbers.

\subsection{Defects4J budget tiers}
Budget fractions are 0.05, 0.10, 0.20, and 0.40 of measured full-suite candidate cost. Native timing and deterministic Python quality are separate. Math and Chart use 21 timing blocks with 200 inner repetitions and 20 warmups; Time and Mockito use the same 21-block design with 10 inner repetitions and 2 warmups. The prepared-mask descriptor audit and budget-sweep Pareto signatures are included as machine-readable CSV files in the accompanying artifact package.

\subsection{Separated quality/timing experiments}
Epinions and BWSN TTD use independent reseeded quality trials and separate paired timing blocks. Timing gathered during parallel quality throughput is excluded from publication timing.

\subsection{Persistent reuse}
The repeated Math experiment uses horizons 1, 5, 10, 20, and 100, active fraction 0.9, five sequences, seven outer runs, and warmup 2. Equality of coverage, selected candidates, and active sequence is checked for each comparison.

\section{Embedded Realization Details}
\label{app:hardware}
The embedded system comprises PicoRV32 RV32IM, 64 KiB instruction ROM, 64 KiB data RAM, a memory-mapped CAGE accelerator, a 64-bit SoC cycle counter, and a simulation mailbox. Accelerated operations include marginal gain, combined gain-and-update, subset, covered-state load/read, and population count. The accelerator maintains persistent covered state. These details document the simulation architecture only; synthesis-specific metrics are intentionally absent.

\section*{Statements and Declarations}

\paragraph*{Funding}
The author received no specific funding for this work.

\paragraph*{Competing interests}
The author declares no competing interests that are relevant to the content of this article.

\paragraph*{Author contributions}
Amirreza Khorasanian conceived the study, developed the methodology and software, designed and conducted the experiments, analyzed and interpreted the results, prepared the visualizations, and wrote and revised the manuscript.

\paragraph*{Data availability}
Processed benchmark inputs, audited result summaries, experiment manifests, and measurement-level data supporting the figures and tables are archived in the CAGE research artifact at \url{https://doi.org/10.5281/zenodo.22114770}. Third-party source datasets and benchmark projects remain subject to their original terms; the artifact provides provenance and regeneration tooling where redistribution is intentionally avoided.

\paragraph*{Code availability}
The CAGE implementation, experiment runners, preparation scripts, and reproduction documentation are available in the \href{https://github.com/amirz81/cage-coverage-optimization}{CAGE GitHub repository} and are archived as release \textbf{v1.0.0} at \url{https://doi.org/10.5281/zenodo.22114770} \cite{cage_artifact2026}. CAGE-owned source code and documentation are distributed under the \textbf{MIT License}; third-party benchmark material remains subject to its original terms.

\bibliographystyle{unsrt}
\bibliography{references}

\begin{thebibliography}{10}

\bibitem{rothermel1999}
Gregg Rothermel, Roland~H. Untch, Chengyun Chu, and Mary~Jean Harrold.
\newblock Test case prioritization: An empirical study.
\newblock In {\em Proceedings of the IEEE International Conference on Software
  Maintenance}, pages 179--188, 1999.

\bibitem{rothermel2001}
Gregg Rothermel, Roland~H. Untch, Chengyun Chu, and Mary~Jean Harrold.
\newblock Prioritizing test cases for regression testing.
\newblock {\em IEEE Transactions on Software Engineering}, 27(10):929--948,
  2001.

\bibitem{elbaum2001}
Sebastian Elbaum, Alexey~G. Malishevsky, and Gregg Rothermel.
\newblock Incorporating varying test costs and fault severities into test case
  prioritization.
\newblock In {\em Proceedings of the 23rd International Conference on Software
  Engineering}, pages 329--338, 2001.

\bibitem{leskovec2007}
Jure Leskovec, Andreas Krause, Carlos Guestrin, Christos Faloutsos, Jeanne
  VanBriesen, and Natalie Glance.
\newblock Cost-effective outbreak detection in networks.
\newblock In {\em Proceedings of the 13th ACM SIGKDD International Conference
  on Knowledge Discovery and Data Mining}, pages 420--429. ACM, 2007.

\bibitem{ostfeld2008}
Avi Ostfeld, James~G. Uber, Elad Salomons, Jonathan~W. Berry, William~E. Hart,
  Cindy~A. Phillips, Jean-Paul Watson, et~al.
\newblock The battle of the water sensor networks ({BWSN}): A design challenge
  for engineers and algorithms.
\newblock {\em Journal of Water Resources Planning and Management},
  134(6):556--568, 2008.

\bibitem{nemhauser1978}
George~L. Nemhauser, Laurence~A. Wolsey, and Marshall~L. Fisher.
\newblock An analysis of approximations for maximizing submodular set functions
  -- i.
\newblock {\em Mathematical Programming}, 14(1):265--294, 1978.

\bibitem{feige1998}
Uriel Feige.
\newblock A threshold of ln n for approximating set cover.
\newblock {\em Journal of the ACM}, 45(4):634--652, 1998.

\bibitem{khuller1999}
Samir Khuller, Anna Moss, and Joseph Naor.
\newblock The budgeted maximum coverage problem.
\newblock {\em Information Processing Letters}, 70(1):39--45, 1999.

\bibitem{sviridenko2004}
Maxim Sviridenko.
\newblock A note on maximizing a submodular set function subject to a knapsack
  constraint.
\newblock {\em Operations Research Letters}, 32(1):41--43, 2004.

\bibitem{minoux1978}
Michel Minoux.
\newblock Accelerated greedy algorithms for maximizing submodular set
  functions.
\newblock In {\em Optimization Techniques}, pages 234--243. Springer, 1978.

\bibitem{badanidiyuru2014}
Ashwinkumar Badanidiyuru and Jan Vondr{\'a}k.
\newblock Fast algorithms for maximizing submodular functions.
\newblock In {\em Proceedings of the Twenty-Fifth Annual ACM-SIAM Symposium on
  Discrete Algorithms}, pages 1497--1514. SIAM, 2014.

\bibitem{mirzasoleiman2015}
Baharan Mirzasoleiman, Ashwinkumar Badanidiyuru, Amin Karbasi, Jan Vondr{\'a}k,
  and Andreas Krause.
\newblock Lazier than lazy greedy.
\newblock In {\em Proceedings of the Twenty-Ninth AAAI Conference on Artificial
  Intelligence}, pages 1812--1818, 2015.

\bibitem{rice1976}
John~R. Rice.
\newblock The algorithm selection problem.
\newblock In {\em Advances in Computers}, volume~15, pages 65--118. Elsevier,
  1976.

\bibitem{xu2008}
Lin Xu, Frank Hutter, Holger~H. Hoos, and Kevin Leyton-Brown.
\newblock {SATzilla}: Portfolio-based algorithm selection for {SAT}.
\newblock {\em Journal of Artificial Intelligence Research}, 32:565--606, 2008.

\bibitem{kotthoff2014}
Lars Kotthoff.
\newblock Algorithm selection for combinatorial search problems: A survey.
\newblock {\em AI Magazine}, 35(3):48--60, 2014.

\bibitem{just2014}
Ren{\'e} Just, Darioush Jalali, and Michael~D. Ernst.
\newblock {Defects4J}: A database of existing faults to enable controlled
  testing studies for java programs.
\newblock In {\em Proceedings of the 23rd International Symposium on Software
  Testing and Analysis}, pages 437--440. ACM, 2014.

\bibitem{richardson2003}
Matthew Richardson, Rakesh Agrawal, and Pedro Domingos.
\newblock Trust management for the semantic web.
\newblock In {\em The Semantic Web -- ISWC 2003}, volume 2870 of {\em Lecture
  Notes in Computer Science}, pages 351--368. Springer, 2003.

\bibitem{chambi2016}
Samy Chambi, Daniel Lemire, Owen Kaser, and Robert Godin.
\newblock Better bitmap performance with roaring bitmaps.
\newblock {\em Software: Practice and Experience}, 46(5):709--719, 2016.

\bibitem{cage_artifact2026}
Amirreza Khorasanian.
\newblock Cage: Regime-aware algorithm engineering for coverage optimization --
  research artifact.
\newblock Zenodo, \url{https://doi.org/10.5281/zenodo.22114770}, 2026.
\newblock Version 1.0.0; MIT License for CAGE-owned code and documentation.

\end{thebibliography}
\end{document}